# Renormalization Group Functions of the $\varphi^4$ Theory in the Strong Coupling Limit: Analytical Results

## I. M. Suslov

*Kapitza Institute for Physical Problems, Russian Academy of Sciences, ul. Kosygina 2, Moscow, 119334 Russia*
*e-mail: suslov@kapitza.ras.ru*

**Abstract**—The previous attempts of reconstructing the Gell-Mann–Low function $\beta(g)$ of the $\varphi^4$ theory by summing perturbation series give the asymptotic behavior $\beta(g) = \beta_\infty g^\alpha$ in the limit $g \longrightarrow \infty$, where $\alpha \approx 1$ for the space dimensions $d = 2,3,4$. It can be hypothesized that the asymptotic behavior is $\beta(g) \sim g$ for all values of $d$. The consideration of the zero-dimensional case supports this hypothesis and reveals the mechanism of its appearance: it is associated with a zero of one of the functional integrals. The generalization of the analysis confirms the asymptotic behavior $\beta(g) = \beta_\infty g$ in the general $d$-dimensional case. The asymptotic behaviors of other renormalization group functions are constant. The connection with the zero-charge problem and triviality of the $\varphi^4$ theory is discussed.

## 1. INTRODUCTION

According to [1], the relation between the bare charge $g_0$ and the observed charge $g$ in the renormalizable field theories has the form

$$g = \frac{g_0}{1 + \beta_2 g_0 \ln(\Lambda/m)}, \quad (1)$$

where $m$ is the particle mass and $\Lambda$ is the momentum cutoff. The zero-charge situation ($g \longrightarrow 0$) takes place at finite $g_0$ and $\Lambda \longrightarrow \infty$. The correct interpretation of formula (1) consists in its inversion

$$g_0 = \frac{g}{1 - \beta_2 g \ln(\Lambda/m)}, \quad (2)$$

so that $g_0$ refers to a distance scale of $\Lambda^{-1}$ and is chosen to fit the observed charge $g$. As $\Lambda$ increases, $g_0$ also increases and Eqs. (1, 2) become inapplicable, so that the existence of the so-called Landau pole in Eq. (2) has no physical sense.

The real behavior of the charge $g(L)$ as a function of the distance scale $L$ is specified by the Gell-Mann–Low equation

$$-\frac{dg}{d\ln L} = \beta(g) = \beta_2 g^2 + \beta_3 g^3 + \ldots \quad (3)$$

and depends on the form of the function $\beta(g)$. According to the Bogoliubov–Shirkov classification [2], the increase in $g(L)$ ceases if $\beta(g)$ has a zero at finite $g$ value and continues infinitely if $\beta(g)$ is nonalternating and has the asymptotic behavior $\beta(g) \sim g^\alpha$ with $\alpha \leq 1$ in the limit $g \longrightarrow \infty$; if $\beta(g) \sim g^\alpha$ with $\alpha > 1$, then $g(L) \longrightarrow \infty$ at finite $L = L_0$ (a real Landau pole appears) and the theory is inconsistent due to indeterminacy of $g(L)$ at $L < L_0$. Landau and Pomeranchuk [3] attempted to justify the last possibility arguing that formula (1) is valid with no restrictions; however, this is possible only under the strict equality $\beta(g) = \beta_2 g^2$, which is invalid because $\beta_3$ is finite.

The above consideration shows that the solution of the zero charge problem requires the determination of the form of the Gell-Mann–Low function $\beta(g)$ at arbitrary $g$ and, in particular, its asymptotic behavior at $g \longrightarrow \infty$. Such an attempt was made in the recent works for the $\varphi^4$ theory [4], QED [5], and QCD [6] (see also review [7]). It was based on the fact that the first four coefficients $\beta_N$ in Eq. (3) are known from diagrammatic calculations [8, 10, 11], whereas the asymptotic behavior for large $N$ values has the form $\beta_N^{as} = c a^N \Gamma(\gamma N + b)$, which is calculated using the Lipatov method [7, 12–15]. The corrections to the asymptotic expression has the form of the regular expansion in $1/N$

$$\beta_N = \beta_N^{as}\left\{1 + \frac{A_1}{N} + \frac{A_2}{N^2} + \ldots + \frac{A_K}{N^K} + \ldots\right\}, \quad (4)$$

which allows the interpolation of the coefficient function by truncating the series and choosing the coefficients $A_K$ from the correspondence with the known val-ues $\beta_2$, $\beta_3$, $\beta_4$, and $\beta_5$. For the variation of the interpolation procedure, series (4) can be re-expanded as

$$\beta_N = \beta_N^{as}\left\{1 + \frac{\tilde{A}_1}{N - \tilde{N}} + \frac{\tilde{A}_2}{(N - \tilde{N})^2} + \ldots \right. \\ \left. + \frac{\tilde{A}_K}{(N - \tilde{N})^K} + \ldots \right\}, \quad (5)$$

by introducing an arbitrary parameter $\tilde{N}$. The summation of the series for the four-dimensional $\varphi^4$ theory [4] provides non-alternating $\beta$ function and the results for the exponent $\alpha$ are independent of $\tilde{N}$ within the error (see Fig. 1), indicating that $\alpha$ is close to 1. Similar results for three- and two-dimensional $\varphi^4$ theories (see Fig. 2) were obtained recently in [19, 20], where critical exponents were calculated. It can be hypothesized that the asymptotic behavior is linear, $\beta(g) \sim g$, for an arbitrary space dimension $d$. The simplicity of the result indicates that it can be obtained analytically.

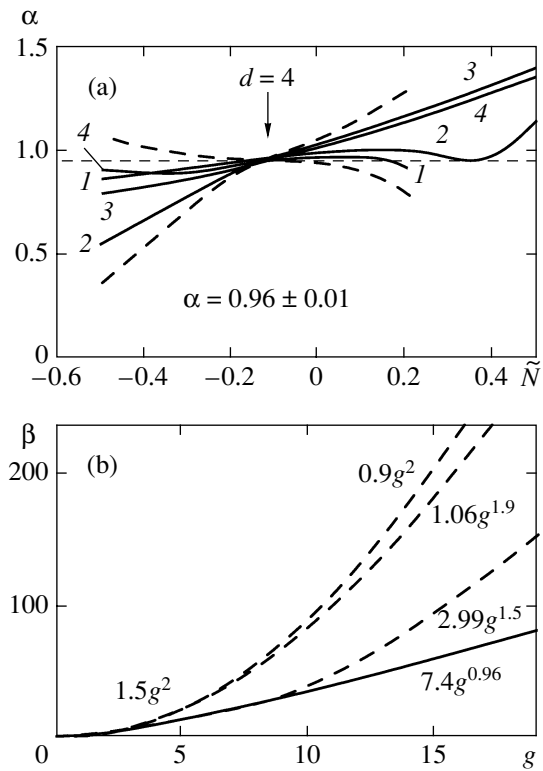

**Fig. 1.** Results for the four-dimensional $\varphi^4$ theory: (a) exponent $\alpha$ (its various estimates were described in [4]) versus $\tilde{N}$ and (b) the general form of the Gell-Mann–Low function according to (solid line) [4] and (dashed lines from top to bottom) [16–18].

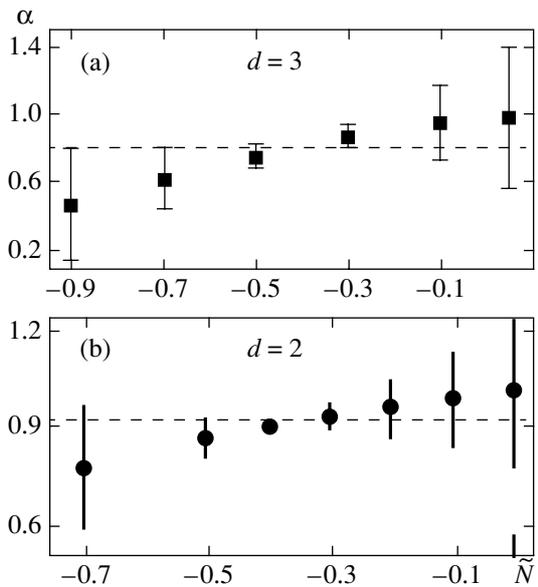

**Fig. 2.** Results for the exponent $\alpha$ in the $\varphi^4$ theory for the space dimensions $d = 3$ [20] and 2 [19].

It will be shown below that this is so indeed. The analysis of the zero-dimensional case (Sect. 3) confirms the existence of the linear asymptotic behavior $\beta(g) \sim g$ and reveals the mechanism of its appearance. It is associated with a surprising circumstance that the limit $g \longrightarrow \infty$ for the renormalized charge $g$ is determined not by large values of the bare charge $g_0$ (as seems to be intuitively obvious), but by its complex values. Moreover, it is possible to analyze only the region $|g_0| \ll 1$, where the functional integrals can be estimated in the saddle-point approximation. If the direction in the complex plane $g_0$ is chosen so that the contribution from the trivial vacuum is comparable with the saddle-point contribution from the leading instanton, the functional integral can vanish. The limit $g \longrightarrow \infty$ is associated with the zero of one of the functional integrals; as a result, this limit is controllable and allows one to obtain the asymptotic behaviors of the $\beta$ function and anomalous dimensions (Sect. 2): the first asymptotic behavior is indeed linear in the general $d$-dimensional case (Sect. 4) in the reasonable agreement with the summation results (Sect. 5).

In the four-dimensional case, the asymptotic result $\beta(g) = \beta_\infty g$ in combination with the definiteness of the sign of the $\beta$ function (see Fig. 1b) corresponds to the second possibility in the Bogoliubov–Shirkov classification: the effective interaction is finite at long distances $L \gtrsim m^{-1}$, but unboundedly increases as $g(L) \sim$ $g(L) \sim L^{-\beta_\infty}$ at $L \longrightarrow 0$. This contradicts the commonly accepted opinion that the continuum $\varphi^4$ theory is trivial. In fact, two different definitions of triviality appear to be confused in the literature (see Sect. 6). The first definition introduced by Wilson [21] is equivalent to the positiveness of $\beta(g)$ at $g \neq 0$, which is confirmed by entire available information and can be considered as a firmly established property. The second definition appearing in mathematical works [22–24] corresponds to the true triviality and is equivalent to the internal inconsistency according to Bogoliubov and Shirkov: it requires not only the positive sign of the β function, but also its sufficiently fast increase at infinity. The indications to the true triviality are scarce and allow another interpretation (see Sect. 6). This analysis provides a new view on this problem: to arrive at a nontrivial theory, it is necessary to use complex values of the bare charge $g_0$, which have been never considered neither in mathematical proofs nor in numerical experiments.

## 2. DEFINITION OF THE RENORMALIZATION GROUP FUNCTIONS

Below, the $n$ component $\varphi^4$ theory is considered with the action

$$S\{\varphi\} = \int d^d x \left\{ \frac{1}{2} \sum_\alpha (\nabla \varphi_\alpha)^2 + \frac{1}{2} m_0^2 \sum_\alpha \varphi_\alpha^2 + \frac{1}{8} u \left( \sum_\alpha \varphi_\alpha^2 \right)^2 \right\}, \quad (6)$$

$$u = g_0 \Lambda^\epsilon, \quad \epsilon = 4 - d$$

in the $d$-dimensional space, where $g_0$ and $m_0$ are the bare charge and mass, respectively. The most general functional integral in this theory contains $M$ factors of the field $\varphi$ in the pre-exponential:

$$Z^{(M)}_{\alpha_1 \ldots \alpha_M}(x_1, \ldots, x_M)$$
$$= \int D\varphi \, \varphi_{\alpha_1}(x_1) \varphi_{\alpha_2}(x_2) \ldots \varphi_{\alpha_M}(x_M) \exp(-S\{\varphi\}), \quad (7)$$

and is related to the $M$-point Green's functions $G^{(M)} = Z^{(M)}/Z^{(0)}$. These Green's functions make it possible to determine the "amputated" vertices $\Gamma^{(L, N)}$ with $N$ external lines of the field $\varphi$ and $L$ external interaction lines[1]; the simplest vertices are shown in Fig. 3. The multiplicative renormalizability of the vertex $\Gamma^{(L, N)}$ means that [26][2]

$$\Gamma^{(L, N)}(p_i; g_0, m_0, \Lambda)$$
$$= Z^{-N/2} \left( \frac{Z_2}{Z} \right)^{-L} \Gamma_R^{(L, N)}(p_i; g, m), \quad (8)$$

where $p_i$ are the external momenta; i.e., the divergence at $\Lambda \to \infty$ disappears after the extraction of the corresponding $Z$ factors and transformation to the renormal-

---
[1] In terms of the diagrammatic technique presented in [25], where the interaction is denoted by the dotted lines.

[2] The $\varphi^4$ theory is not renormalizable at $d > 4$ and the subsequent consideration is meaningful only for $d \leq 4$.

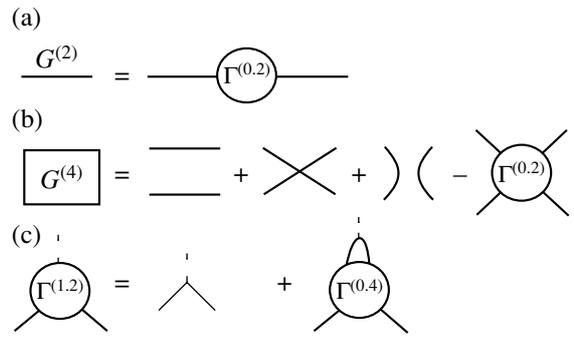

**Fig. 3.** Relations between the "amputated" vertices $\Gamma^{(L, N)}$ and the Green's functions $G^{(M)}$.

ized charge ($g$) and mass ($m$). The renormalization conditions at zero momentum are accepte

$$\Gamma_R^{(0, 2)}(p; g, m)\big|_{p \to 0} = m^2 + p^2 + O(p^4),$$
$$\Gamma_R^{(0, 4)}(p_i; g, m)\big|_{p_i = 0} = g m^\epsilon, \quad (9)$$
$$\Gamma_R^{(1, 2)}(p_i; g, m)\big|_{p_i = 0} = 1,$$

which are usually used in the theory of phase transitions [27]. The substitution of Eq. (8) into Eqs. (9) expresses $g$, $m$, $Z$, and $Z_2$ in terms of the bare quantities:

$$Z(g_0, m_0, \Lambda) = \left( \frac{\partial}{\partial p^2} \Gamma^{(0, 2)}(p; g_0, m_0, \Lambda)\big|_{p = 0} \right)^{-1},$$
$$Z_2(g_0, m_0, \Lambda) = \left( \Gamma^{(1, 2)}(p_i; g_0, m_0, \Lambda)\big|_{p_i = 0} \right)^{-1}, \quad (10)$$
$$m^2 = Z(g_0, m_0, \Lambda) \Gamma^{(0, 2)}(p; g_0, m_0, \Lambda)\big|_{p = 0},$$
$$g m^\epsilon = Z^2(g_0, m_0, \Lambda) \Gamma^{(0, 4)}(p_i; g_0, m_0, \Lambda)\big|_{p_i = 0}.$$

The Callan–Symanzik equation is obtained by application of the differential operator $d/d\ln m$ on Eq. (8) at fixed $g_0$ and $\Lambda$ values [26]:

$$\left[ \frac{\partial}{\partial \ln m} + \beta(g) \frac{\partial}{\partial g} + (L - N/2) \eta(g) - L \eta_2(g) \right] \quad (11)$$
$$\times \Gamma^{(L, N)} \approx 0,$$

and is valid asymptotically at large $p_i/m$ values. The renormalization group functions $\beta(g)$ (Gell-Mann–Low function) and $\eta(g)$ and $\eta_2(g)$ (anomalous dimensions) are defined as

$$\beta(g) = \frac{dg}{d\ln m}\bigg|_{g_0, \Lambda = \text{const}},$$
$$\eta(g) = \frac{d\ln Z}{d\ln m}\bigg|_{g_0, \Lambda = \text{const}}, \quad (12)$$
$$\eta_2(g) = \frac{d\ln Z_2}{d\ln m}\bigg|_{g_0, \Lambda = \text{const}}$$

and generally depend on all variables; in fact, they depend only on $g$ according to general theorems [26].

## 3. ZERO-DIMENSIONAL CASE

### 3.1. "Naive" Zero-Dimensional Limit

For the passage to the zero-dimensional limit, let us consider the system bounded in all spatial directions at a sufficiently small scale; in this case, the coordinate dependence $\varphi(x)$ can be neglected and the terms with the gradients can be omitted in Eq. (7). Treating the functional integral as a multiple integral on a lattice and choosing the sufficiently sparse lattice, one can think that only one site of the lattice locates inside the system. In this case,

$$Z^{(M)}_{\alpha_1 \ldots \alpha_M} = \int d^n \varphi \, \varphi_{\alpha_1} \ldots \varphi_{\alpha_M} \exp\left(-\frac{1}{2}m_0^2 \varphi^2 - \frac{1}{8}u\varphi^4\right). \quad (13)$$

The diagrammatic technique generated by "functional integrals" (13) has the usual form, but all propagators are taken at zero momentum and the integration with respect to the momenta is absent. In a fixed order of perturbation theory, all diagrams are equal to each other and their total contribution is determined by the combinatorics of the diagrams; conversely, this combinatorics can be studied by means of functional integrals (13) [29].

The above notions on the zero-dimensional theory are commonly accepted. However, they do not quite correspond to the correct zero-dimensional limit of the $\varphi^4$ theory. Considering the simplest diagrams, it is easy to verify (see Appendix 1) that the indicated trivialization of the diagrammatic technique occurs only on zero external momenta; if these momenta are nonzero, no noticeable simplifications appear in the zero-dimensional limit. The last circumstance is significant for the determination of the $Z$ factor, which is introduced through the scheme

$$G^{(2)}(p) = (p^2 + m_0^2 + \Sigma(p, m_0))^{-1} = (p^2 + m_0^2$$
$$+ a_0(m_0) + a_2(m_0)p^2 + a_4(m_0)p^4 + \ldots)^{-1} \quad (14)$$
$$= \frac{Z}{p^2 + m^2 + O(p^4)},$$

i.e., is determined by the momentum dependence of the self-energy [cf. Eq. (10)]. In the above naive zero-dimensional theory, the momentum dependence is absent and does not require a special normalization; for this reason, $Z = 1$ is taken. Such a procedure is self-consistent, but it does not correspond to the correct zero-dimensional limit of the $\varphi^4$ theory. The last circumstance is insignificant, because the described model is used only for illustration: below, the general $d$-dimensional case will be considered.

### 3.2. General Expressions for the Renormalization Group Functions

The substitution of $\varphi_\alpha = \varphi u_\alpha$ into Eq. (13) and the introduction of the integration with respect to the directions of the unit vector $\mathbf{u}$ provide

$$Z^{(M)}_{\alpha_1 \ldots \alpha_M} = \int_0^\infty \varphi^{M+n-1} d\varphi \exp\left(-\frac{1}{2}m_0^2 \varphi^2 - \frac{1}{8}u\varphi^4\right) \quad (15)$$
$$\times \int d^n u \, \delta(|u|-1) u_{\alpha_1} \ldots u_{\alpha_M}.$$

This expression after the calculation of the integral with respect to $d^n u$ [28] for even $M$ values reduces to the form

$$Z^{(M)}_{\alpha_1 \ldots \alpha_M} = \frac{2\pi^{n/2}}{2^{M/2} \Gamma(M/2 + n/2)} \quad (16)$$
$$\times I_{\alpha_1 \ldots \alpha_M} K_M(m_0, u),$$

where $I_{\alpha_1 \ldots \alpha_M}$ is the sum of the terms $\delta_{\alpha_1 \alpha_2} \delta_{\alpha_3 \alpha_4} \ldots$ with all possible pairings and

$$K_M(m_0, u) = \int_0^\infty \varphi^{M+n-1} d\varphi \quad (17)$$
$$\times \exp\left(-\frac{1}{2}m_0^2 \varphi^2 - \frac{1}{8}u\varphi^4\right).$$

The separation of the factor $I_{\alpha_1 \ldots \alpha_M}$ in the Green's functions and vertices:

$$G^{(2)}_{\alpha\beta} = G_2 \delta_{\alpha\beta}, \quad G^{(4)}_{\alpha\beta\gamma\delta} = G_4 I_{\alpha\beta\gamma\delta}, \quad (18)$$
$$\Gamma^{(0,2)}_{\alpha\beta} = \Gamma_2 \delta_{\alpha\beta}, \quad \Gamma^{(0,4)}_{\alpha\beta\gamma\delta} = \Gamma_4 I_{\alpha\beta\gamma\delta},$$

provides

$$\Gamma_2 = 1/G_2, \quad G_4 = G_2^2 - G_2^4 \Gamma_4, \quad (19)$$

where

$$G_2 = \frac{1}{n} \frac{K_2(m_0, u)}{K_0(m_0, u)},$$
$$G_4 = \frac{1}{n(n+2)} \frac{K_4(m_0, u)}{K_0(m_0, u)} \quad (20)$$

and the vertex $\Gamma^{(0,4)}_{\alpha\beta\gamma\delta}$ is determined by the usual relation (see Fig. 3b)

$$G^{(4)}_{\alpha\beta\gamma\delta} = G^{(2)}_{\alpha\beta} G^{(2)}_{\gamma\delta} + G^{(2)}_{\alpha\gamma} G^{(2)}_{\beta\delta} + G^{(2)}_{\alpha\delta} G^{(2)}_{\beta\gamma} \quad (21)$$
$$- G^{(2)}_{\alpha\alpha'} G^{(2)}_{\beta\beta'} G^{(2)}_{\gamma\gamma'} G^{(2)}_{\delta\delta'} \Gamma^{(0,4)}_{\alpha'\beta'\gamma'\delta'}.$$

According to renormalization conditions (10),

$$m^2 = \Gamma_2 = \frac{nK_0}{K_2}, \quad (22)$$

$$g = \frac{\Gamma_4}{m^4} = 1 - m^4 G_4 = 1 - \frac{n}{n+2}\frac{K_4 K_0}{K_2^2}. \quad (23)$$

Since the derivative of $K_M$ reduces to $K_{M+2}$ (see Eq. (17)), the differentiation of Eq. (22) with respect to $m_0^2$ gives

$$\frac{dm^2}{dm_0^2} = \frac{n}{2}\left\{-1 + \frac{K_4 K_0}{K_2^2}\right\}. \quad (24)$$

Since all differentiations in Eq. (12) are performed at $g_0, \Lambda =$ const, it is convenient to treat these parameters as fixed; then, $m^2$ is a function of only $m_0^2$ and formula (24) can be "inverted", i.e., treated as an expression for the derivative $dm_0^2/m^2$. According to the definition of the β function in Eqs. (12),

$$\beta(g) = 2\frac{dg}{d\ln m^2}$$
$$= -\frac{2m^4}{n(n+2)}\left[2\frac{K_4}{K_0} + \left(\frac{K_4}{K_0}\right)'_{m_0^2} m^2 \frac{dm_0^2}{dm^2}\right]. \quad (25)$$

In view of Eq. (24), this expression gives

$$\beta(g) = -\frac{2n}{n+2}\frac{K_4 K_0}{K_2^2}\left[2 + \frac{K_6 K_0/K_4 K_2 - 1}{1 - K_4 K_0/K_2^2}\right]. \quad (26)$$

Change $\varphi \longrightarrow \varphi(8/u)^{1/4}$ reduces integrals (17) to the form

$$K_M(t) = \int_0^\infty \varphi^{M+n-1} d\varphi \exp(-t\varphi^2 - \varphi^4),$$
$$t = \left(\frac{2}{u}\right)^{1/2} m_0^2. \quad (27)$$

The appearing factors drop out of the combinations $K_4 K_0/K_2^2$ and $K_6 K_0/K_4 K_2$ on which Eqs. (23) and (26) depend; correspondingly, these expressions hold their forms when the integrals $K_M(m_0, u)$ are changed to the integrals $K_M(t)$. The right-hand sides of Eqs. (23) and (26) are functions of one variable $t$, and the $\beta(g)$ dependence is determined by these formulas in the parametric form.

The vertex $\Gamma_{\alpha\beta}^{(1,2)} = \Gamma_{12}\delta_{\alpha\beta}$ is determined by the Ward identity [61]

$$\Gamma_{12} = \frac{dm^2}{dm_0^2} = 1 - \frac{n+2}{2}g, \quad (28)$$

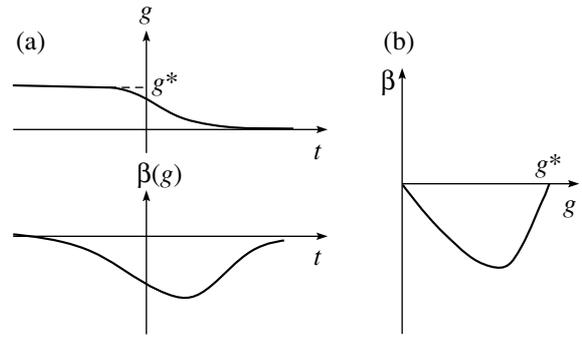

**Fig. 4.** (a) Qualitative behavior of $g$ and $\beta(g)$ with variation of $t$ along the real axis and (b) corresponding $\beta(g)$ dependence.

which makes it possible to obtain the following expression for $\eta_2(g)$:

$$\eta_2(g) = -\frac{d\ln\Gamma_{12}}{d\ln m} = \frac{\beta(g)}{2/(n+2) - g}. \quad (29)$$

In the accepted approximation, the function $\eta(g)$ is identically zero.

### 3.3. Analysis of the Renormalization Group Functions

Using the asymptotic expressions

$$K_M(t) =$$
$$\begin{cases} \frac{1}{\sqrt{2}} t^{-(M+n)/2} \Gamma\left(\frac{M+n}{2}\right) \\ \times \left[1 - \frac{(M+n)(M+n+2)}{4 \times t^2} + \ldots\right], \\ t \longrightarrow \infty, \\ \frac{1}{4}\left[\Gamma\left(\frac{M+n}{4}\right) - t\Gamma\left(\frac{M+n+2}{4}\right) + \ldots\right], \\ t \longrightarrow 0, \\ \frac{\sqrt{\pi}}{2} e^{t^2/4} \left(\frac{|t|}{2}\right)^{(M+n-2)/2} \\ \times \left[1 + \frac{(M+n-2)(M+n-4)}{4t^2} + \ldots\right], \\ t \longrightarrow -\infty, \end{cases} \quad (30)$$

it is easy to verify that the dependence of $g$ and $\beta(g)$ on $t$ has the form shown in Fig. 4a; i.e., variation of the parameter $t$ along the real axis governs the behavior of $\beta(g)$ from zero to the fixed point (see Fig. 4b)[3]

$$g^* = \frac{2}{n+2}. \quad (31)$$

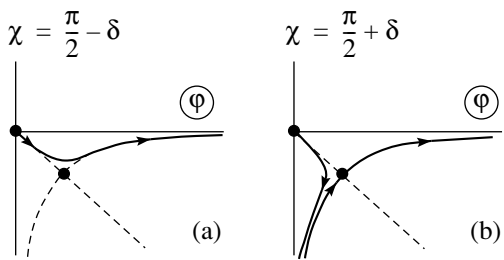

**Fig. 5.** Topology of the steepest descend lines for the integral $K_M(t)$ for various values of $\chi = \arg t$: the steepest descend line for (a) $0 < |\chi| < \pi/2$ passes only through the trivial saddle point, whereas both saddle points are passed for (b) $\pi/2 < |\chi| < \pi$.

To pass to larger $g$ values, it is necessary to analyze the parametric representation specified by Eqs. (23) and (26) at complex $t$ values. Let $t = |t|e^{i\chi}$ and $|t| \gg 1$; then, depending on the phase $\chi$, the integrals $K_M(t)$ are determined either by the trivial saddle point $\varphi = 0$ or by the nontrivial saddle point $\varphi^2 = -t/2$. The saddle-point values of the integrals $K_M(t)$ depend on $\chi$, but this dependence cancels in the combinations $K_4 K_0 / K_2^2$ and $K_6 K_0 / K_4 K_2$ determining Eqs. (23) and (26). For this reason, in the rough approximation, the complex plane $t$ is separated into two regions, where $g$ and $\beta(g)$ are constant: $g = 0$, $\beta(g) = 0$ and $g = g^*$, $\beta(g) = 0$. There is a smooth transition between these values; it is associated with the deviations from the saddle-point approximation, which appear for $|t| \ll 1$. However, the expected changes occur in finite limits as for real $t$ values (see Fig. 4a). It is easy to understand that large $g$ values can be reached only near the directions in the complex plane $t$ for which the contributions from two saddle points are comparable. In this case, $K_M(t)$ is represented as

$$K_M(t) = Ae^{i\psi} + A_1 e^{i\psi_1} = Ae^{i\psi}(1 + ke^{i\Delta}) \qquad (32)$$

and one can attempt to turn the integral to zero by varying the parameters $k$ and $\Delta$. The existing degrees of freedom are sufficient for this, because the real and imaginary parts of $t$ can be varied. When $t$ is varied, the coefficient $k$ surely passes through unity, because there are regions in the complex plane $t$ where one of two terms in Eq. (32) dominates. The phase $\Delta$ varies in the infinite limits (see below), so that the number of the zeros of the integral $K_M(t)$ is infinite. They locate along the rays $\chi = \pm 3\pi/4$, concentrating at infinity; this reasoning is strictly justified for zeros locating in the region $|t| \gg 1$, where the saddle-point approximation is applicable.

---

[3] The presence of the fixed point $g^*$ does not mean the existence of the phase transition, which is absent at $d < 2$. Indeed, the Callan–Symanzik equation determining the scaling behavior of the correlation functions is valid only for small values of $m$, which cannot be reached with physical $m_0$ and $g_0$ values. Formula (31) is in agreement with the result $\tilde{g}^* = (n + 8)/(n + 2)$ obtained in [31], where the definition of the charge $\tilde{g}$ differs from the definition accepted in this work by a factor, $\tilde{g} = (n + 8)g/2$. This result does not correspond to the correct zero-dimensional limit of $\varphi^4$ theory and its use in the interpolation scheme refining the dependence of $g^*$ on the space dimension $d$ [31] is not reasonable.

It is easy to see that the limit $g \longrightarrow \infty$ can be reached by tending $K_2$ to zero; in this case, Eqs. (23) and (26) are simplified:

$$g \approx -\frac{n}{n+2}\frac{K_4 K_0}{K_2^2}, \quad \beta(g) \approx -\frac{4n}{n+2}\frac{K_4 K_0}{K_2^2}, \qquad (33)$$

and the parametric representation is resolved in the form

$$\beta(g) = 4g, \quad g \longrightarrow \infty, \qquad (34)$$

whereas it follows from Eq. (29) that

$$\eta_2(g) = -4, \quad g \longrightarrow \infty. \qquad (35)$$

In agreement with expectation, the asymptotic behavior of the $\beta$ function appears to be linear.

### 3.4. Zeros of the Integrals $K_M(t)$

When deriving results (34) and (35), the explicit form of the integrals $K_M(t)$ is not used: it is substantial only that they can vanish in principle and that the zeros of different integrals $K_M(t)$ locate at different points. Let us show that these assumptions are justified.

The action at the saddle points $\varphi = 0$ and $\varphi^2 = -t/2$ is $0$ and $t^2/4$, respectively, and the contributions from two saddle points are comparable for $\mathrm{Re}\, t^2 = 0$ or $\chi = \pm\pi/4$, $\pm 3\pi/4$. However, the values $\chi = \pm\pi/4$ are inappropriate. The integral $K_M(t)$ exhibits the Stokes phenomenon associated with the change in the topology of the steepest descend lines [32]; it occurs at $|\chi| = \pi/2$, so that the steepest descend line in the region $0 < |\chi| < \pi/2$ passes only through the trivial saddle point (see Fig. 5a), whereas both fixed points are passed for $\pi/2 < |\chi| < \pi$ (see Fig. 5b). Therefore, the compensation of saddle-point contributions (32) is possible at $\chi = \pm 3\pi/4$, but it does not occur at $\chi = \pm\pi/4$. With $t = \rho e^{i\chi}$, $\rho \gg 1$ and $\chi = 3\pi/4 + \Delta$, $\Delta \ll 1$, the contribution from two saddle points to the integral $K_0(t)$ is expressed as

$$K_0(t) = \rho^{-n/2}\exp\left(-i\frac{3\pi}{8}n\right)\left[\frac{1}{2}\Gamma\left(\frac{n}{2}\right) + \frac{\sqrt{\pi}}{2^{n/2}}\right. \\
\left. \times \exp\left(-i\frac{\pi}{4} + i\frac{\pi}{4}n - i\frac{1}{4}\rho^2\right)\rho^{n-1}\exp\left(\frac{1}{2}\rho^2\Delta\right)\right]. \qquad (36)$$

This expression with $\Delta(\rho)$ chosen from the condition

$$\rho^{n-1}\exp\left(\frac{1}{2}\rho^2\Delta\right) = \frac{2^{n/2-1}}{\sqrt{\pi}}\Gamma\left(\frac{n}{2}\right), \text{ i.e.}$$

$$\Delta \sim \frac{\ln\rho}{\rho^2}, \quad (37)$$

gives

$$K_0(t) = \frac{1}{2}\Gamma\left(\frac{n}{2}\right)\rho^{-n/2}\exp\left(-i\frac{3\pi}{8}n\right)$$
$$\times\left[1+\exp\left(\frac{i}{4}(\pi+\pi n\rho^2)\right)\right] \quad (38)$$

and the zeros of the integral $K_0(t)$ appear at the points

$$\rho_s^2 = \pi(n+5)+8\pi s, \quad s\text{—integer.} \quad (39)$$

The results for $K_M(t)$ are obtained by changing $n \longrightarrow n+M$ and it is clear from Eqs. (37) and (39) that different integrals $K_M(t)$ vanish at different points.

Another method for obtaining the zeros of the integrals $K_M(t)$ involves the use of special functions. The simplest integral of the zero-dimensional $\varphi^4$ theory is related to the second-kind modified Bessel function $\mathcal{K}_\nu(x)$ as

$$F(g) = \int_{-\infty}^{\infty} d\phi \exp(-\phi^2 - g\phi^4)$$
$$= \frac{1}{2}g^{-1/2}e^{1/8g}\mathcal{K}_{1/4}\left(\frac{1}{8g}\right). \quad (40)$$

This relation is easily derived noting that $F(g)$ satisfies the equation [33]

$$4g^2 F'' + (8g+1)F' + \frac{3}{4}F = 0 \quad (41)$$

with the boundary condition $F(0) = \sqrt{\pi}$. Hence, at $n = 0$,

$$K_0(t) = \int_0^\infty d\phi\, e^{-t\phi^2-\phi^4}$$
$$= \frac{1}{4}t^{1/2}e^{t^2/8}\mathcal{K}_{1/4}\left(\frac{t^2}{8}\right). \quad (42)$$

The second-kind modified Bessel function $\mathcal{K}_\nu(x)$ has no zeros on the principal sheet of the Riemann surface ($|\arg z|<\pi$), but has zeros on the neighboring sheets. At large $|z|$ values, they have the form[4]

$$z_s = -\frac{1}{2}\ln(2\cos\pi\nu)+e^{\pm 3\pi i/2}\left(\frac{3\pi}{4}+\pi s\right), \quad (43)$$
$s$—integer.

It is clear from Eqs. (42) and (43) that $K_0(t)$ has zeros at the points

$$t_s^2 = -2\ln 2 - 6\pi i + 8\pi s e^{3\pi i/2} \quad (|t|\gg 1), \quad (44)$$

in agreement with Eq. (39) at $n = 0$. The results for $K_M(t)$ with $M = 2, 4, \ldots$ can be obtained by differentiating Eq. (41) with respect to $t$, and their analytic continuation to noninteger $M$ values and change $M \longrightarrow M + n$ provide the generalization to the case $n \neq 1$.

## 4. GENERAL $d$-DIMENSIONAL CASE

### 4.1. Expressions for the Renormalization Group Functions

Since complex $t$ values in the limit $|t| \longrightarrow \infty$ correspond to complex $g_0$ values in the limit $|g_0| \longrightarrow 0$ (see Eq. (27)), the above analysis provides a surprising conclusion: large values of the renormalized charge $g$ correspond not to large values of the bare charge $g_0$ (as naturally expected)[5], but to its complex values; moreover, it is sufficient to analyze only the region $|g_0| \ll 1$, where the saddle-point method is applicable. The above analysis is based only on, first, the possibility of the expression of the renormalization group functions in terms of the functional integrals and, second, the possibility of the investigation of functional integrals in the saddle-point approximation: both foundations can be generalized to an arbitrary $d$-dimensional case.

Let us introduce the Fourier transforms of integrals (7):

$$Z^{(M)}_{\alpha_1\ldots\alpha_M}(p_1, \ldots, p_M)\mathcal{N}\delta_{p_1+\ldots+p_M}$$
$$= \sum_{x_1,\ldots,x_M} Z^{(M)}_{\alpha_1\ldots\alpha_M}(x_1,\ldots,x_M) \quad (45)$$
$$\times e^{ip_1 x_1+\ldots+ip_M x_M},$$

---

[4] Using the known relation for the Airy function $Ai(x) \sim \mathcal{K}_{1/3}\left(\frac{2}{3}x^{2/3}\right)$ and noting that $Ai(x)$ has zeros at negative $x$ values, it is easy to believe that this result is valid.

[5] It is usually assumed that it is possible to introduce a universal function $g = f(L)$ describing the charge as a function of the distance scale; in this case, the observed charge corresponds to $g_{\text{obs}} = f(m^{-1})$, the bare charge corresponds to $g_0 = f(\Lambda^{-1})$, and the renormalized charge at the scale $L$ is $g = f(L)$; i.e., all charges appearing in the theory are the same charge, but refer to different scales. In fact, it is well-known that this picture is approximate due to the ambiguity of the renormalization scheme. The definitions of the bare and renormalized charges are technically different and are introduced in the cutoff and subtraction schemes, respectively [34]. The corresponding functions $g_0 = f_1(L)$ and $g = f_2(L)$ coincide with each other only at the one- and two-loop levels, but are different in higher loops. For this reason, the indicated intuitive concepts are based on the experience in the weak coupling region.

where $N$ is the number of the sites of the lattice on which the functional integral is defined. The choice of the momenta corresponding to the so-called symmetric point, $p_i p_j = p^2(4\delta_{ij} - 1)/3$, allows one to separate the $\delta$ factors from $Z^{(M)}$ as in Eq. (16):

$$Z^{(0)} = K_0, \quad Z^{(2)}_{\alpha\beta}(p, -p) = K_2(p)\delta_{\alpha\beta},$$
$$Z^{(4)}_{\alpha\beta\gamma\delta}\{p_i\} = K_4\{p_i\}I_{\alpha\beta\gamma\delta}. \tag{46}$$

Let us introduce the vertex $\Gamma^{(0,4)}$ through the relation (see Fig. 3b)

$$G^{(4)}_{\alpha\beta\gamma\delta}(p_1, ..., p_4) = G^{(2)}_{\alpha\beta}(p_1)G^{(2)}_{\gamma\delta}(p_3)\mathcal{N}\delta_{p_1+p_2}$$
$$+ G^{(2)}_{\alpha\gamma}(p_1)G^{(2)}_{\beta\delta}(p_2)\mathcal{N}\delta_{p_1+p_3}$$
$$+ G^{(2)}_{\alpha\delta}(p_1)G^{(2)}_{\beta\gamma}(p_3)\mathcal{N}\delta_{p_1+p_4} - G^{(2)}_{\alpha\alpha'}(p_1)G^{(2)}_{\beta\beta'}(p_2)$$
$$\times G^{(2)}_{\gamma\gamma'}(p_3)G^{(2)}_{\delta\delta'}(p_4)\Gamma^{(0,4)}_{\alpha'\beta'\gamma'\delta'}(p_1, ..., p_4) \tag{47}$$

and separate the $\delta$ factors as in Eq. (46)

$$G^{(2)}_{\alpha\beta}(p, -p) = G_2(p)\delta_{\alpha\beta},$$
$$G^{(4)}_{\alpha\beta\gamma\delta}\{p_i\} = G_4\{p_i\}I_{\alpha\beta\gamma\delta}, \tag{48}$$
$$\Gamma^{(0,4)}_{\alpha\beta\gamma\delta}\{p_i\} = \Gamma_4\{p_i\}I_{\alpha\beta\gamma\delta}.$$

It is inconvenient to take $p_i = 0$, because the relation between $G_4$ and $\Gamma_4$ contains the factors $\mathcal{N}$ proportional to the volume. It is more convenient to take $p_i \sim \mu$, excluding the special equalities such as $p_1 + p_2 = 0$, and then to take $\mu$ so that $\mathcal{L}^{-1} \lesssim \mu \ll m$, where the lower boundary tends to zero in the limit of the infinite size of the system $\mathcal{L}$. In this case,

$$G_4 = \frac{K_4}{K_0}, \quad \Gamma_4 = -\frac{G_4}{G_2^4} = -\frac{K_4 K_0^3}{K_2^4}, \tag{49}$$

where the integrals are taken at zero momenta and

$$G_2 = \frac{K_2(p)}{K_0},$$
$$\Gamma_2(p) = \frac{1}{G_2(p)} = \frac{K_0}{K_2(p)} \approx \frac{K_0}{K_2} + \frac{K_0\tilde{K}_2}{K_2^2}p^2, \tag{50}$$

where the following expansion for small $p$ values is used:

$$K_2(p) = K_2 - \tilde{K}_2 p^2 + ... \tag{51}$$

Then, the $Z$ factors, renormalized mass, and renormalized charge are given by the expressions

$$Z = \left[\frac{\partial}{\partial p^2}\Gamma_2(p)\right]^{-1}_{p=0} = \frac{K_2^2}{K_0\tilde{K}_2}, \tag{52}$$

$$m^2 = Z\Gamma_2(p=0) = \frac{K_2}{\tilde{K}_2}, \tag{53}$$

$$g = m^{(-\epsilon)}Z^2\Gamma_4 = -\left(\frac{K_2}{\tilde{K}_2}\right)^{d/2}\frac{K_4 K_0}{K_2^2}, \tag{54}$$

$$\frac{1}{Z2} = \Gamma_{12}\{p=0\} = \frac{d(1/G2(0))}{dm_0^2}, \quad \frac{dm^2}{dm_0^2} =$$
$$= \left(\frac{K_2}{\tilde{K}_2}\right)' = \frac{K_2'\tilde{K}_2 - K_2\tilde{K}_2'}{\tilde{K}_2^2}, \tag{55}$$

where primes mean the derivatives with respect to $m_0^2$. As in Section 3, it is convenient to consider the parameters $g_0$ and $\Lambda$ as fixed; in this case, $m^2$ is a function of $m_0^2$ and the derivative $dm_0^2/dm^2$ is determined by the expression inverse to Eq. (55). According to the definitions of the renormalization group functions given by Eqs. (12),

$$\beta(g) = \frac{dg}{d\ln m} = -dm^d\frac{K_4 K_0}{K_2^2}$$
$$- 2m^{d+2}\left(\frac{K_4 K_0}{K_2^2}\right)'_{m_0^2}\frac{dm_0^2}{dm^2},$$
$$\eta(g) = \frac{d\ln Z}{d\ln m} \tag{56}$$
$$= 2m^2[\ln K_2^2 - \ln K_0 - \ln\tilde{K}_2]'_{m_0^2}\frac{dm_0^2}{dm^2},$$
$$\eta_2(g) = \frac{d\ln Z2}{d\ln m} = -2m^2\left[\ln\frac{K_0'K - K_0K_2'}{\tilde{K}_2^2}\right]'_{m_0^2}\frac{dm_0^2}{dm^2}.$$

In view of Eq. (55), these expressions reduce to the form

$$\beta(g) = \left(\frac{K_2}{\tilde{K}_2}\right)^{d/2}\left\{-d\frac{K_4 K_0}{K_2^2}\right.$$
$$+ 2\frac{(K_4'K_0 + K_4 K_0')K_2 - 2K_4 K_0 K_2'}{K_2^2} \tag{57}$$
$$\left.\times \frac{\tilde{K}_2}{K_2\tilde{K}_2' - K_2'\tilde{K}_2}\right\},$$

$$\eta(g) = -\frac{2K_2\tilde{K}_2}{K_2\tilde{K}_2' - K_2'\tilde{K}_2}\left[2\frac{K_2'}{K_2} - \frac{K_0'}{K_0} - \frac{\tilde{K}_2'}{\tilde{K}_2}\right], \tag{58}$$

$$\eta_2(g) = \frac{2K_2\tilde{K}_2}{K_2\tilde{K}_2' - K_2'\tilde{K}_2}$$
$$\times \left\{ \frac{K_0K_2'' - K_0''K_2}{K_0K_2' - K_0'K_2} - 2\frac{K_2'}{K_2} \right\}. \tag{59}$$

Expressions (54), (57)–(59) specify $\beta(g)$, $\eta(g)$, and $\eta_2(g)$ in the parametric form: the right-hand sides of these formulas with fixed $g_0$ and $\Lambda$ values are functions only of $m_0^2$, whereas the dependence on a particular choice of $g_0$ and $\Lambda$ is absent according to the general theorems (Sect. 2).

### 4.2. Asymptotic Expressions for the Renormalization Group Functions

According to Eq. (54), the limit $g \longrightarrow \infty$ can be reached in two ways: by tending $K_2$ or $\tilde{K}_2$ to zero. In the limit $\tilde{K}_2 \longrightarrow 0$,

$$\beta(g) = -d\left(\frac{K_2}{\tilde{K}_2}\right)^{d/2}\frac{K_4K_0}{K_2^2}, \tag{60}$$
$$\eta(g) \longrightarrow 2, \quad \eta_2(g) \longrightarrow 0$$

and the parametric representation is resolved in the form

$$\beta(g) = dg, \quad \eta(g) = 2,$$
$$\eta_2(g) = 0 \quad (g \longrightarrow \infty) \tag{61}$$

In the case $K_2 \longrightarrow 0$, the limit $g \longrightarrow \infty$ is reached only for $d < 4$:

$$\beta(g) = (d-4)g, \quad \eta(g) = 4,$$
$$\eta_2(g) \longrightarrow 4, \quad (g \longrightarrow \infty). \tag{62}$$

Results (61) and (62) likely correspond to two branches of the function $\beta(g)$. It is easy to understand that the first branch is physical. According to the modern concepts, the properties of the $\varphi^4$ theory change smoothly with change in the space dimension, and the results for $d = 2, 3$ can be obtained by analytic continuation from the dimension $d = 4 - \epsilon$. All available information indicates that $\beta(g)$ is non-alternating at $d = 4$ (Sect. 6), so that its asymptotic expression for $g \longrightarrow \infty$ is positive. According to continuity, the positive asymptotic expression is expected for $d < 4$. Result (61) has such a property, whereas the region of large $g$ values for branch (62) cannot be reached at $d = 4$. The approximate results for $\beta(g)$ mentioned in Section 1 also indicate that result (61) is valid. Finally, expression (61) at $d = 2$ is in agreement with the asymptotics $\beta(g) = 2g$ for the Ising model obtained from the duality relation [35].

The above consideration implies that the mechanism of the appearance of the asymptotic behavior of the renormalization group functions is the same as in the naive zero-dimensional limit. Strictly speaking, the possibility of reaching large $g$ values due to another mechanism, e.g., due to a large $K_4$ value, is not excluded. However, this possibility seems to be low probable: considering the field $\varphi(x)$ localized at the unit scale and estimating the pre-exponential factor in Eq. (7) for a typical field configuration, we obtain $K_M \sim \langle\varphi\rangle^M K_0$, $\tilde{K}_2 \sim K_2$, and the substitution into Eq. (54) provides $g \sim 1$. Change in the common scale of all lengths does not affect $g$, due to its dimensionless nature. Therefore, it is impossible to obtain large $g$ values due to change in the amplitude of the field $\varphi(x)$ or in the common scale of its spatial localization. In any case, it should be assumed that the mean value $\langle\varphi\rangle$ is anomalously small for one of the integrals due to some reason (e.g., because the function $\varphi(x)$ is alternating); however, this assumption returns us to the variants considered above.

### 4.3. Zeros of the Functional Integrals

For complex $g_0$ values such that $|g_0| \ll 1$, the zeros of the functional integrals can be obtained from the condition of the compensation of the contribution from trivial vacuum and the saddle-point contribution from the instanton configuration with the minimum action.[6] The latter contribution is well studied and has the form (see, e.g., [36])

$$[Z^{(M)}_{\alpha_1...\alpha_M}(p_1, ..., p_M)]^{\text{inst}} = ic_M(-g_0)^{-(M+r)/2}$$
$$\times e^{-S_0/g_0}\langle\phi_c\rangle_{p_1}...\langle\phi_c\rangle_{p_M}I_{\alpha_1...\alpha_M} \tag{63}$$

for $d < 4$ and a somewhat more complicated form at $d = 4$. Here, $\langle\phi_c\rangle_p$ is Fourier transform of the dimensionless instanton configuration $\phi_c(x)$, $S_0$ is the corresponding action, $r$ is the number of the zero modes ($r = n + d - 1$ for $d < 4$ and $r = n + 4$ for $d = 4$), and $c_M$ is a constant. Then, for $M = 0, 2, \ldots$,

$$Z_0 = 1 + ic_0(-g_0)^{-r/2}e^{-S_0/g_0},$$
$$Z^{(2)}_{\alpha\beta}(p, p') = \frac{\delta_{\alpha\beta}}{p^2 + m_0^2} \tag{64}$$
$$+ ic_2(-g_0)^{-(r+2)/2}e^{-S_0/g_0}\langle\phi_c\rangle^2\delta_{\alpha\beta},$$

etc. The substitution of $t^2 = -S_0/g_0$ provides the expression similar to Eq. (36), which are analyzed similarly. It is easy to verify that the zeros of different integrals $K_M$ and their derivatives with respect to $m_0^2$ are different.

---

[6] All instanton singularities for the $\varphi^4$ theory in the Borel plane lie on the negative semiaxis [7]; for this reason, the action for all instantons can be considered as positive at the appropriate choice of the complex phase $g_0$

Comparison with the summation results

|  | $d = 2$ | $d = 3$ | $d = 4$ |
|---|---|---|---|
| Eq. (61) | $\alpha = 1$<br>$\beta_\infty = 2$ | $\alpha = 1$<br>$\beta_\infty = 3$ | $\alpha = 1$<br>$\beta_\infty = 4$ |
| Summation of the series [4, 19, 20] | $\alpha = 0.92 \pm 0.02$<br>$\beta_\infty = 22 \pm 3$ | $\alpha = 0.84 \pm 0.07$<br>$\beta_\infty = 60 \pm 10$ | $\alpha = 0.96 \pm 0.01$<br>$\beta_\infty = 14.8 \pm 0.8$ |
| Summation at $\alpha = 1$ | $\beta_\infty = 10.6$ | $\beta_\infty = 16.8$ | $\beta_\infty = 10.6$ |

It is easy to show that the effect of higher instantons is insignificant near the zero of the integral $\tilde{K}_2$, where it is obvious that $e^{-S_0/g_0} \sim |g_0|^{(r+2)/2}$. The higher instantons can be classified as follows.

(a) *Combinations of k remote elementary instantons.* For them, the number of the zero modes is $r_k = kr$ and action is $S_k = kS_0$; as a result, the following extra factor appears as compared to Eq. (63)

$$[(-g_0)^{-r/2} e^{-S_0/g_0}]^{k-1} \sim |g_0|^{k-1}. \quad (65)$$

(b) *Higher spherically symmetric instantons.* They have the same symmetry and the same number of zero modes $r$ as the leading instanton, but their action $\tilde{S}$ is larger. Their contribution differs from contribution (63) in the extra factor

$$e^{-(\tilde{S}-S_0)/g_0} \sim |g_0|^{(r+2)(\tilde{S}-S_0)/2S_0}, \quad (66)$$

which is small in the case of interest $r + 2 > 0$.

(c) *Localized asymmetric instantons.* They have larger action $S_{as}$ and larger number of zero modes $r_{as} = r + d(d-1)/2$ due to the possibility of rotation in the coordinate space [36]. Their contribution contains the extra factor

$$(-g_0)^{-d(d-1)/4} e^{-(S_{as}-S_0)/g_0}$$
$$\sim |g_0|^{-d(d-1)/4 + (r+2)(S_{as}-S_0)/2S_0}. \quad (67)$$

For known asymmetric instantons, the ratio $S_{as}/S_0$ is rather large (see discussion in [36]) and the exponent in Eq. (67) is positive.

(d) *Combinations of several remote instantons of types (b) and (c).* As is easily verified, their contribution contains an additional smallness as compared to Eqs. (66) and (67).

## 5. REMARKS ON THE SUMMATION RESULTS

The summation of the perturbation series makes it possible to obtain the asymptotic behavior $\beta_\infty g^\alpha$ with the exponent $\alpha$ close to unity for the $\beta$ function (see Sect. 1) in agreement with result (61). The results for the coefficient $\beta_\infty$ [4, 19, 20] are given in the table and are in much worse agreement with Eq. (61).[7]

This poor agreement is not surprising, because currently acquired information indicates an insufficiently reliable estimate of $\beta_\infty$. In particular, a test experiment on the contraction of information of the $\varphi^4$ theory was performed in [5]. The complete information includes the coefficients $\beta_2$, $\beta_3$, $\beta_4$, $\beta_5$, parameters $a$, $b$, $c$, and $\gamma$ of the Lipatov asymptotic expression (Sect. 1), and coefficient $A_1$ in Eq. (4); its contraction provides the following.

*Complete information:*

$$\alpha = 0.96 \pm 0.01, \quad \beta_\infty = 14.8 \pm 0.8.$$

*Excluding $A_1$:*

$$\alpha = 1.00 \pm 0.01, \quad \beta_\infty = 6.8 \pm 0.6.$$

*Excluding $A_1$ and $c$:*

$$\alpha = 1.02 \pm 0.03, \quad \beta_\infty = 3.4 \pm 0.6.$$

Even a more efficient test experiment was obtained for QED due to the error, when the Lipatov asymptotic expression in the summation of the series in [5] was taken with an extra factor of $(4\pi)^N$:

*Correct asymptotic expression:*

$$\alpha = 1.0 \pm 0.1, \quad \beta_\infty = 1.0 \pm 0.3.$$

*Extra factor $(4\pi)^N$:*

$$\alpha = 1.0 \pm 0.2, \quad \beta_\infty = -3 \times 10^3.$$

It is easy to see that the estimates of the exponent $\alpha$ exhibit high stability, whereas the coefficient $\beta_\infty$ is quite

---

[7] Values for $\beta_\infty$ at $d = 4$ differ by a factor of 2 from values in [4, 5], because Eq. (3) in those works was written with $L^2$ instead of $L$. The summation results for $d = 4$ refer to another renormalization scheme (MOM), but this is likely insignificant. The renormalized charge in physical renormalization schemes is determined from the same vertex $\Gamma_4$, but with different its relations to the distance scale $L$. For a power-law $g(L)$ dependence, this difference can provide only a constant factor, so that the definitions of the charge in different schemes coincide in the order of magnitude; hence, the asymptotic behavior $\beta_\infty g$ must be identical in them ($g(L) \propto L^{-\beta_\infty}$ at small distances and the difference in $\beta_\infty$ would lead to an arbitrarily large difference between the charges). The formal results for the MOM scheme do not contradict this analysis, but do not confirm it more clearly (see Appendix II).

sensitive to the amount and quality of available information.

If the value $\alpha = 1$ is treated as known when summing the series, the results for $\beta_\infty$ significantly shift towards the correct values (see the last row of table)[8]; their high sensitivity even to small errors in $\alpha$ is seen. It is associated with the obvious fact that, when determining $\beta_\infty$, any uncertainties in $\alpha$ are multiplied by a large factor and appear in the exponential.

The asymptotic behavior $\sim g^2$ for the function $\eta(g)$ was obtained in [19, 20]; this result seemed to significantly contradict Eq. (61). However, the separation $\eta(g) = \eta_2 g^2 + \tilde{\eta}(g)$ was performed in [19, 20], and the asymptotic behavior $\tilde{\eta}(g) = Ag^2$ with the correct exponent was obtained. According to Eq. (61), the coefficient $A$ must be $-\eta_2$, but the accuracy of the numerical procedure was too low to reveal this.[9] The situation is similar for $\eta_2(g)$ for which the asymptotic behavior $\sim g$ was obtained in [19, 20].

## 6. IS THE $\varphi^4$ THEORY TRIVIAL?

In the four-dimensional case, result (61) for the asymptotic behavior of the $\beta$ function in combination with the fact that this function is positive (see Fig. 1b) means the realization of the second possibility in the Bogoliubov–Shirkov classification (see Sect. 1): the effective interaction is finite at large distances, but increases infinitely at small $L$ values as $g(L) \sim L^{-4}$. This conclusion contradicts the commonly accepted conception that the continual $\varphi^4$ theory is trivial [22–24, 38–60]. It is strange that this conception is commonly accepted, because real attempts of investigating the strong-coupling region are scarce and their results cannot be considered as established. As shown below, this is explained by the fact that two different definitions of triviality are confused in the literature.

### 6.1. Wilson Triviality

Formula (1) has absolutely another interpretation in the theory of phase transitions. In this case, the cutoff parameter $\Lambda$ and bare charge $g_0$ have direct physical sense and are related to the lattice constant and the coefficient in the effective Landau Hamiltonian. The "zero charge" is reached in the limit $m \longrightarrow 0$ (which corresponds to approaching the phase transition point) and means the absence of the interaction between large-scale fluctuations of the order parameter. This interaction in the $d = 4 - \epsilon$ dimension appears to be finite, but weak in a measure of $\epsilon$, which ensures the success of the Wilson $\epsilon$ expansion [21].

---

[8] The estimate was obtained by shifting the parameter $b_0$ used in [4, 19, 20] from the first minimum until the exact value $\alpha = 1$ is reached: the optimum value for the parameter $\tilde{N}$ was used; the error of the results was not analyzed.

[9] The separation of the term $\eta_2 g^2$ was motivated by two facts. First, a sharp spike appears in the interval $2 < N < 3$ in the interpolation with the use of all coefficients; this spike was interpreted as indication of a singularity. Second, the results for the asymptotic behavior in this case appear to be very indefinite. It is now clear that these facts require another interpretation.

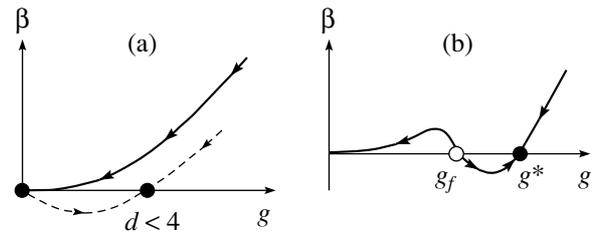

**Fig. 6.** Change in $g$ when the Gell-Mann–Low equation is integrated towards large $L$ values: (a) evolution for a non-alternating function $\beta(g)$ finishes at the Gaussian fixed point $g = 0$; (b) for an alternating function $\beta(g)$, the boundary $g_f$ of the attraction region of the Gaussian fixed point appears. The function $\beta(g)$ at $d < 4$ has a negative portion [dashed line in panel (a)].

In more recent works, Wilson passed to a deeper formulation of the question: Is the indicated triviality of the four-dimensional theory characteristic of small $g_0$ values or of the global character? The answer to this question is determined by the properties of the $\beta$ function: if $\beta(g)$ has no nontrivial zeros (see Fig. 6a), the effective interaction vanishes in the limit of large distances irrespectively of the initial $g_0$ value. If $\beta(g)$ is alternating (see Fig. 6b), a nontrivial limit $g^*$ can appear at large distances. The latter possibility is of great interest for condensed matter physics: this is the problem of the existence of the new-type phase transitions to which the Wilson $\epsilon$ expansion is inapplicable [61].

Using reductio ad absurdum, Wilson assumed the existence of the boundary $g_f$ of the attraction region of the Gaussian fixed point $g = 0$ (which is equivalent to the property that $\beta(g)$ is alternating) and derived the consequences from this assumption that are convenient for numerical verification. According to his results [21], no indications of the existence of the point $g_f$ are revealed. This was the first attempt of investigating the strong-coupling region for the $\varphi^4$ theory and the first evidence that $\beta(g)$ is non-alternating.

### 6.2. Mathematical Triviality

Another definition of triviality was proposed in mathematical works [22–24]. If field theory is treated as the limit of lattice theories, the bare charge $g_0$ can be considered as a function of the interatomic distance $a_0$. If it is possible to pass to the limit $a_0 \longrightarrow 0$ with a certain choice of the function $g_0(a_0)$ and to ensure a finite interaction at large distances, the theory is nontrivial; if this is impossible with any choice of $g_0(a_0)$, the theory is trivial. This definition corresponds to the notion of true triviality, i.e., the fundamental impossibility of the development of the continual theory with a finite interaction at large distances; this is equivalent to Bogoliubov–Shirkov internal inconsistency (see Sect. 1). Indeed, in the last case, owing to the finiteness of the charge $g_\infty$ at large distances, the theory does not exist at scales $L < L_0$; the passage to the limit $a_0 \longrightarrow 0$ requires a decrease in $L_0$ to zero, which is possible only at $g_\infty \longrightarrow 0$.

The triviality of the $\varphi^4$ theory at $d > 4$ and its nontriviality at $d < 4$ were strictly proven in [22–24], and nonrigourous reasoning in favor of the triviality at $d = 4$ was given using the experience of those proofs. These results are physically quite obvious. Indeed, the nonrenormalizability of the $\varphi^4$ theory at $d > 4$ means that the limit $a_0 \longrightarrow 0$ is impossible without the destruction of the theory structure; since the structure of the $\varphi^4$ theory in the accepted definition of the triviality, is supported artificially at arbitrarily small $a_0$ values, the only possibility for it is to "throw off" the interaction and to pass to the Gaussian theory. The nontriviality of the $\varphi^4$ theory at $d < 4$ is associated with the existence of the negative portion for the $\beta$ function (dashed line in Fig. 6a), which can be verified for $d = 4 - \epsilon$ with a small $\epsilon$ value and be numerically confirmed for $d = 2$ and 3: it is easy to see that $g(L) \longrightarrow g^*$ at large distances and $g(L) \longrightarrow 0$ at small distances for this portion.

The above discussion clearly indicates that the results proved in [22–24] do not require analysis of the strong-coupling region, and they do not provide foundation for any conclusions on the situation at $d = 4$, where such an analysis is necessary. Finally, note that complex values of the bare charge were not considered in the mathematical works, but the consideration of such values is necessary for developing the nontrivial theory at $d = 4$.

The above consideration clarifies the difference between two definitions of triviality. For Wilson triviality, it is sufficient for $\beta(g)$ to have a definite sign, whereas true triviality additionally requires its sufficiently fast increase $\beta(g) \sim g^\alpha$ with $\alpha > 1$ in the strong-coupling region. However, this difference is practically not realized in the literature. The authors of some works (see, e.g., [40, 47]) directly state that the limits $\Lambda \longrightarrow \infty$ and $m \longrightarrow 0$ are equivalent. The formal solution of Eq. (3),

$$\int_{g_m}^{g_\Lambda} \frac{dg}{\beta(g)} = \ln\frac{\Lambda}{m} \qquad (68)$$

is indeed determined only by the ratio $\Lambda/m$; however, its physical consequences depend on the formulation of the problem. If $\Lambda$ and $g_\Lambda$ are fixed, $g_m \longrightarrow 0$ at $m \longrightarrow 0$ for $\beta(g) > 0$. If $m$ and $g_m$ are fixed, $g_\Lambda \longrightarrow \infty$ together with $\Lambda \longrightarrow \infty$ is possible only at $\alpha \leq 1$; otherwise, the limit $\Lambda \longrightarrow \infty$ is impossible.

### 6.3. Specificity of the β Function at d = 4

The general form of the $\beta$ function for the four-dimensional $\varphi^4$ theory that was obtained in [4] by summing the perturbation series is shown in Fig. 1b along with the results obtained by other authors.[10] The positive sign of $\beta(g)$ and, hence, the Wilson triviality are doubtless. There are also reasons to expect manifestations of true triviality. Note that the "natural" charge normalization is used in Fig1b, for which the parameter $a$ of the Lipatov asymptotics is unity; this normalization corresponds to the interaction term in the form $(16\pi^2/4!)g\varphi^4$. In this case, the nearest singularity of the Borel transform is at the unit distance from the coordinate origin; therefore, $\beta(g)$ changes at a scale of about unity. Nevertheless, the applicability region of the single-loop law appears to be somewhat extended and the behavior close to square continues to $g \sim 10$. In the traditional charge normalizations, such extension is even longer, to $g \sim 10^3$ when the interaction term is written in the form $g\varphi^4/8$ or $g\varphi^4/4!$ Taking into account that the $\beta$ function is convex up to $g \sim 100$ (in the natural normalization) [4], it is clear that the behavior of any quantities is indistinguishable from the trivial behavior in a wide region of the parameter values.

### 6.4. Numerical Results

The existing numerical results can be separated into several groups.

(a) *Decrease in g(L) with an increase in L.* The decrease in the effective interaction $g(L)$ was obtained in many works (see, e.g., [38–40]) and indicates only that $\beta(g)$ is positive. Detailed analysis of this decrease allows one in principle to acquire information on the $\beta$ function, but such an analysis has never been performed.

(b) *Renormalization group in real space.* This is approximate implementation of the Kadanoff construction [25] following the early works by Wilson. The procedure is based on reduction of the description by division of the system into the blocks and their subsequent consolidation; the blocks are characterized by a finite number of the parameters whose evolution is then observed. The works in this direction are of high quality [41, 42], but they demonstrate only the evolution of the system to the Gaussian fixed point and confirm the initial Wilson analysis.

(c) *Logarithmic corrections to scaling.* Phase transitions at $d > 4$ are described by the mean field theory, whereas the corresponding power laws at $d = 4$ are supplemented by the logarithmic corrections [26, 61]

---

[10] Of course, the particular form of $\beta(g)$ somewhat changes when a correct asymptotic result (61) is used instead of the approximate asymptotics found in [4].

$$M \propto (-\tau)^{1/2}[\ln(-\tau)]^{3/(n+8)},$$
$$\chi^{-1} \propto |\tau|[\ln|\tau|]^{-(n+2)/(n+8)}, \qquad (69)$$
$$H \propto M^3/|\ln M|, \quad \tau = 0,$$

etc., where $M$, $H$, $\chi$, and $\tau$ are the magnetization, magnetic field, susceptibility, and distance to the transition point in temperature. The logarithmic corrections undoubtedly exist and their numerical verification [43–50] confirms either (at $g_0 \ll 1$) the results of the leading logarithmic approximation [61] or (at $g_0 \gtrsim 1$) the Wilson picture of critical phenomena. Nevertheless, most authors directly attribute their results to the triviality of the $\varphi^4$ theory.

(d) *Expansion of Eq. (1) to the region of large $g_0$ values.* In our opinion, the dependence of the renormalized charge on the bare charge at a fixed ratio $\Lambda/m$ that was analyzed in [51–54] is the single evidence of the true triviality of the $\varphi^4$ theory. The characteristic example of such results [51] is shown in Fig. 7; it indicates that a Landau pole exists in the dependence of $g_0$ on $L$.

More careful analysis of the results reveals a typical muddle associated with the charge normalization. Reaching $g_0 \sim 400$, Freedman et al. [51] were sure that they reached the strong-coupling region. In fact (see Sect. 6.3), all results for finite $g_0$ values are in the region of the square law for the $\beta$ function and, hence, do not exhibit significant deviations from Eq. (1) (see direct evidence of this in [52]). Only the points for $g_0 = \infty$ that are obtained by reduction to the Ising model are nontrivial, but this reduction is based on the fact that the empirical dependence $m_0^2 = -\text{const}\, g_0$ (corresponding to the single-loop law) is extrapolated to the region of arbitrarily large $g_0$ values. Since there are no reasons for such an extrapolation, the results for $g_0 = \infty$ are unreliable: without these results, nothing follows from Fig. 7. The $g(g_0)$ dependences similar to those shown in Fig. 7 are also obtained from high-temperature series [54] and strong-coupling lattice expansions [53], but they also involve a doubtful extrapolation based on the assumption that the reduction to the Ising model occurs in the way indicated above.

The serious investigations of such kind must firstly reveal real deviations from Eq. (1) that are associated with the deviation from the quadratic dependence of the $\beta$ function; only analysis of these deviations can provide information on the behavior of $\beta(g)$ in the strong-coupling region.

The approach used in Sections 3 and 4 gives a new view on the results under discussion. Since $g(L)$ increases unboundedly at $L \longrightarrow 0$, the development of the nontrivial continual theory requires the use of the complex values of the bare charge: such a possibility was ignored in [51–54]; hence, the picture obtained in those works (see Fig. 7) proves nothing even if it is literally accepted.

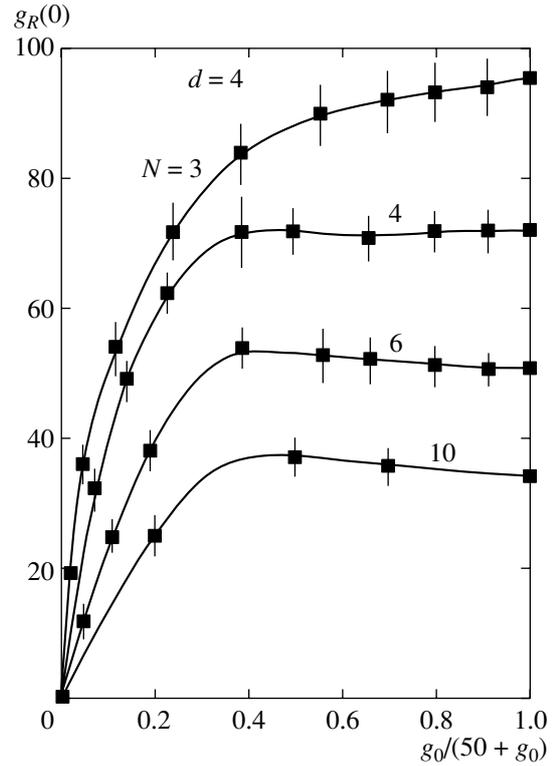

**Fig. 7.** Renormalized charge $g_R(0)$ taken at zero momenta versus the bare charge $g_0$ referring to the interatomic distance $a_0$ in the four-dimensional $\varphi^4$ theory at fixed $Na_0$ and $m$ values with different numbers $N^4$ of the lattice sites (according to [51]).

(d) *Recent works.* The problems associated with triviality have been actively discussed in the recent works by Consoli, Agodi, et al. [55–57], who proposed a nontrivial continual limit of the $\varphi^4$ theory, which leads in fact to the negation of the standard perturbation theory.

The authors illustrate their idea by an example of the imperfect Bose gas with the Bogoliubov spectrum [$\epsilon(k) \sim k$ at small $k$ values and $\epsilon(k) \sim k^2$ at $k \longrightarrow \infty$]. If the passage to the "continual limit" is made by tending two characteristic scales of the problem—the scattering length and distance between the particles–to zero, then either the quadratic spectrum of the ideal gas is recovered ("completely trivial theory") or the strictly linear spectrum of noninteracting phonons appears ("trivial theory with a nontrivial vacuum"), depending on the relation between the scales. The authors proposed the latter scenario for the continual limit of the $\varphi^4$ theory, stating that it is logically consistent.

If the last statement is true, there is an open question of why this passage to the limit physically occurs. In particular, it is impossible to simultaneously vary the gas density and scattering length in the Bose gas of neutral atoms. The situation desired for the authors can appear under the special long-range law: in this case, variation in the density changes the "Debye screening radius," which determines the scattering length; however, such a scenario is not arbitrary and can be predicted on the basis of the initial Hamiltonian.

Consoli, Agodi, et al. [55–57] believe that the assumption of the nontrivial character of the continual limit is confirmed by the lattice numerical simulation. However, this conclusion is based on the interpretation of the "experimental" data rather than on the direct data themselves: the numerical experiments are performed deep in the region of the single-loop law and cannot carry any information on the triviality; so their results (even exotic) must be explained in the weak-coupling theory.

### 6.5. Theoretical Results

(a) *Landau–Pomeranchuk reasoning.* Landau and Pomeranchuk [3] noted that, according to Eq. (1), as $g_0$ increases, the observed charge $g$ reaches a value of $1/(\beta_2 \ln \Lambda/m)$, which is independent of $g_0$. Such a behavior can be obtained by changing $\varphi \longrightarrow \tilde{\varphi} g_0^{-1/4}$ in functional integral (7) and omitting $\varphi^2$ terms in action (6); such a change in Eq. (47) gives $G^{(4)}/[G^{(2)}]^2 = \text{const}(g_0)$ and $\Gamma^{(0,4)}[G^{(2)}]^2 \propto \Gamma^{(0,4)} Z^2 \propto \Gamma_R^{(0,4)} = g = \text{const}(g_0)$. If such a procedure is justified even at $g_0 \ll 1$, this is the more so at $g_0 \gtrsim 1$; hence, formula (1) can be considered as applicable at arbitrary $g_0$ values.

At the qualitative level, this reasoning can be correct[11] for the real $g_0$ values, which are assumed in them. According to Sections 3 and 4, change in $g_0$ along the real axis corresponds to change in $g$ from zero to a finite value of $g_{\max}$. The behavior $g_{\max} \longrightarrow 0$ at $\Lambda \longrightarrow \infty$ would mean the qualitative validity of Eq. (1); the above Monte Carlo results (see Fig. 7) point to this possibility. The development of the nontrivial theory requires the use of complex $g_0$ values with $|g_0| \lesssim 1$; in this case, neither the reduction of the functional integral to the dimensionless form (justified at $|g_0| \gg 1$) nor Eq. (1) is valid; the latter is attributed to the fact that, despite the possibility of using values $|g_0| \ll 1$, perturbation theory is inapplicable, because the instanton contribution is significant.

(b) *Summation of the perturbation series.* The first attempts of the reconstruction of the Gell-Mann–Low function by summing the perturbation series [16–18] led to the asymptotic expression $\beta_\infty g^\alpha$ with $\alpha > 1$, indicating the self-inconsistency (or true triviality) of the $\varphi^4$ theory (see Fig. 1b): this was one of the strongest reasons. The opposite result obtained in [4] at least means that this conclusion does not certainly follow from these investigations.[12] At the same time, all results indicate that $\beta(g)$ is positive and confirm the Wilson triviality.

(c) *Works of the synthetic type.* Works [58] are widely cited as the systematic justification of the triviality of the $\varphi^4$ theory. They constitute an attempt of synthesizing all available information, but contain nothing new for the analysis of the strong-coupling region. Conclusions of works [58] are not surprising, because all easily available information should inevitably indicate triviality due to the specificity of the $\beta$ function discussed in Section 6.3.

(d) *Theories with the $\varphi^p$ interaction.* Some representation on the properties of the $\varphi^4$ theory can be obtained by studying theories with a more general interaction $\varphi^p$. Bender and Jones [59] believe that the consideration of the case $p = 2 + \delta$ with the expansion in the parameter $\delta$ provides serious reasons in a favor of the triviality of the $\varphi^4$ theory. On the other hand, the exact calculation of the $\beta$ function in the limit $p \longrightarrow \infty$ [60] gives the asymptotic behavior $g(\ln g)^{-\gamma}$, which proves the nontriviality of the theory. The second result is more reliable, because it is not associated with the fact that the bare charge is real, as assumed in [59].

(d) *Limit $n \longrightarrow \infty$.* In the limit $n \longrightarrow \infty$, the $\varphi^4$ theory is considered as exactly solvable [25, 62]. In this case, the $\beta$ function is effectively single-loop and leads to the results of type (1), which correspond to the asymptotic behavior $\beta(g) \sim g^2$. This fact is often considered as evidence of the triviality of the $\varphi^4$ theory even in competent works [62].

In fact, the coefficients of the $\beta$ function are polynomials in $n$ and, at $d = 4 - \epsilon$, have the structure

$$\beta(g) = -\epsilon g + \beta_2(n+a)g^2 + \beta_3(n+b)g^3 \\ + \beta_4(n^2 + cn + d)g^4 + \ldots \qquad (70)$$

where $\beta_2, \beta_3, a, \ldots \sim 1$. The change of the variables

$$g = \frac{\tilde{g}}{n}, \quad \beta(g) = \frac{\tilde{\beta}(\tilde{g})}{n} \qquad (71)$$

gives

$$\tilde{\beta}(\tilde{g}) = -\epsilon \tilde{g} + \beta_2 \tilde{g}^2 + \frac{1}{n} f_1(\tilde{g}) + \frac{1}{n^2} f_2(\tilde{g}) + \ldots \qquad (72)$$

---

[11]Their correctness at the quantitative level is excluded, because the $\beta$ function is not quadratic. In fact, the result $g = \text{const}(g_0)$ follows from the reduction of the functional integral to the dimensionless form only at $g_0 \gg 1$, whereas its validity at $g_0 \ll 1$ following from Eq. (1) can be attributed to other causes; this result is likely violated at $g_0 \sim 1$, but the coincidence of the constants in the order of magnitude can be expected from the matching condition.

[12]The results obtained in [16, 17] has an objective origin and are related with the mentioned extension of the single-loop law. They were reproduced in [4] as an intermediate asymptotic behavior and were explained by the characteristic dip in the coefficient function. The variational perturbation theory [18] in the region $g < 10$ gives results close to those obtained in [4], but does not guarantee the correct strong-coupling asymptotic behavior even theoretically.

where only first two terms remain in the limit $n \longrightarrow \infty$. This conclusion is valid for $\tilde{g} \sim 1$ or $g \sim 1/n$, which is sufficient for analyzing $\beta(g)$ near the fixed point and determining the critical exponents. However, such a procedure provides no information on the region $g \sim 1$ and, the more so, on $g \gg 1$. Therefore, no conclusions on the triviality of the $\varphi^4$ theory can be made.

The above analysis shows that the Wilson triviality is confirmed by all available information and can be considered as an established fact. Indications to the true triviality are scarce and allow another interpretation; according to the results of the present work, such a triviality is certainly absent.

## APPENDIX I

### Limit $d \longrightarrow 0$ in the Diagrammatic Technique

Let us consider the simplest integral

$$\Pi(q) = \int \frac{d^d k}{(2\pi)^d} G(k) G(k+q), \tag{A.1}$$

corresponding to the polarization loop. Transforming the propagator according to the scheme [26]

$$G(k) = \frac{1}{k^2 + m^2} = \int_0^\infty da\, e^{-am^2 - ak^2} \tag{A.2}$$

and calculating the appearing Gaussian integral with respect to $k$, we obtain

$$\Pi(q) = \frac{1}{(2\pi)^d} \int_0^\infty da_1 \int_0^\infty da_2 \left(\frac{\pi}{a_1 + a_2}\right)^{d/2} \\ \times \exp\left\{-\frac{a_1 a_2}{a_1 + a_2} q^2 - m^2(a_1 + a_2)\right\}. \tag{A.3}$$

The zero-dimensional limit of this integral is trivially calculated at $q = 0$:

$$\Pi(0) = \frac{1}{m^4} \tag{A.4}$$

and corresponds to the recipe formulated in Section 3: all propagators are taken at zero momenta and the integration with respect to $k$ is absent. At finite $q$ values, the calculation of the integral gives the nontrivial momentum dependence

$$\Pi(q) = \frac{2}{m^2(q^2 + 4m^2)} + \frac{8}{q(q^2 + 4m^2)^{3/2}} \\ \times \ln \frac{\sqrt{q^2 + 4m^2} + q}{2m}, \tag{A.5}$$

which is hardly be obtained for an arbitrary diagram.

In the general case, the expression for the diagram contains $M$ propagators and $L$ integrations with respect to $k_1, \ldots, k_L$. The transformation of the propagators using scheme (A.2) gives the Gaussian integral, which is calculated by the formula [26]

$$\int \prod_{l=1}^L d^d k_l\, e^{-M_{ll'} k_l k_{l'} - 2v_l k_l} \\ = \left(\frac{\pi^L}{\det M}\right)^{d/2} e^{M_{ll'}^{-1} v_l v_{l'}}. \tag{A.6}$$

Quantities $v_l$ are linear in the external momenta; at zero values of these momenta, the zero-dimensional limit of expression (A.6) is unity, and the general expression for the diagram reduces to the integral

$$\int_0^\infty da_1 \ldots \int_0^\infty da_M\, e^{-m^2(a_1 + \ldots + a_M)}, \tag{A.7}$$

which is calculated trivially.

## APPENDIX II

### Other Renormalization Schemes

The so-called MOM scheme, which corresponds to the critical point in the theory of phase transitions, is often used in applications. In this scheme, the value $m_c$ corresponding to zero renormalized mass $m$ is fixed for the bare mass $m_0$. Instead of Eqs. (9), the renormalization conditions are written in the form

$$\Gamma_R^{(0,2)}(p; g, m)\big|_{p^2 = 0} = 0, \\ \frac{\partial}{\partial p^2}\Gamma_R^{(0,2)}(p; g, m)\bigg|_{p^2 = \mu^2} = 1, \\ \Gamma_R^{(0,4)}(p_i; g, m)\big|_{p_i \sim \mu} = g\mu^\epsilon, \\ \Gamma_R^{(1,2)}(p_i; g, m)\big|_{p_i \sim \mu} = 1, \tag{A.8}$$

where $\mu$ is the arbitrary momentum scale.[13] The expressions for the $Z$ factors and renormalized charge in terms of the bare parameters have the form

$$Z = \left(\frac{\partial}{\partial p^2}\Gamma^{(0,2)}(p; g_0, m_c, \Lambda)\bigg|_{p^2 = \mu^2}\right)^{-1}, \\ Z_2 = (\Gamma^{(1,2)}(p_i; g_0, m_0, \Lambda)\big|_{p_i \sim \mu})^{-1}, \\ g = \mu^{-\epsilon} Z^2 \Gamma^{(0,4)}(p_i; g_0, m_0, \Lambda)\big|_{p_i \sim \mu}, \tag{A.9}$$

---

[13]The symmetric point $p_i p_j = \mu^2(4\delta_{ij} - 1)/3$ is usually chosen for $\Gamma^{(0,4)}\{p_i\}$, whereas $p_1^2 = p_2^2 = \mu^2$ and $p_1 p_2 = -\mu^2/3$ are taken for $\Gamma^{(1,2)}(q, p_1, p_2)$

and $m_c$ is determined from the equation $\Gamma^{(0,2)}(0; g_0, m_c, \Lambda) = 0$. With the definition of the β function in the MOM scheme,

$$\beta(g) = \left.\frac{dg}{d\ln\mu}\right|_{g_0, \Lambda = \text{const}}, \quad (A.10)$$

it is easy to obtain the parametric representation

$$g = -\mu^{-\epsilon}\frac{K_4 K_0}{(K'_2)^2}, \quad (A.11)$$

$$\beta(g) = \mu^{-\epsilon}\frac{K_4 K_0}{(K'_2)^2}\left[\epsilon + 4\mu^2\left(\frac{K''_2}{K'_2} - \frac{K'_4}{2K_4}\right)\right], \quad (A.12)$$

where μ is used as a running parameter and the primes mean the derivatives with respect to $\mu^2$. According to Eq. (A.11), the limit $g \longrightarrow \infty$ can be reached in several different ways.

(a) $\mu \longrightarrow 0$ *at finite $K_M$ values and their derivatives.* Then,

$$g = -\mu^{-\epsilon}\frac{K_4 K_0}{(K'_2)^2}, \quad \beta(g) = \epsilon\mu^{-\epsilon}\frac{K_4 K_0}{(K'_2)^2}, \quad (A.13)$$

and the parametric representation is resolved as $\beta(g) = -\epsilon g$, which corresponds to the nonphysical branch in Section 4.

(b) μ = const and $K'_2 \longrightarrow 0$. Excluding $K'_2$, we obtain

$$\beta(g) = 4ig^{3/2}\mu^{2+\epsilon/2}\frac{K''_2}{\sqrt{K_4 K_0}}. \quad (A.14)$$

This formula must be valid at arbitrary ϵ values, because the properties of the $\varphi^4$ theory vary smoothly with $d$. However, the integrals $K_M$ and their derivatives near the zero of $K'_2$ are analytic functions of $\mu^2$ [see Eq. (64)], and the disappearance of the dependence on μ, which is guaranteed by the general theorems, does not occur in Eq. (A.14). Therefore, this variant is self-inconsistent.

(c) $\mu \longrightarrow 0$ and $K'_2 \longrightarrow 0$. In this case,

$$\beta(g) = -g\left[\epsilon + 4\mu^2\frac{K''_2}{K'_2}\right], \quad (A.15)$$

and the dependence on μ can be excluded by setting $K'_2 \longrightarrow 0$. Then,

$$\beta(g) = \text{const}\, g, \quad g \longrightarrow \infty, \quad (A.16)$$

where const must be a number independent of any parameters. This result qualitatively corresponds to Eq. (61), but is less definite.

The presented approach cannot be applied in principle in the minimum subtraction scheme (MS). In this scheme, the definition of the charge does not correspond to the vertex $\Gamma_4$ at a certain choice of the momenta. Therefore, the renormalization group functions cannot be expressed in terms of the functional integrals. As explained in [63], the momentum scale λ for each individual diagram can be taken on the order of μ, so that the usual subtraction at the scale λ is equivalent to the minimum subtraction at the scale μ. However, it is impossible to introduce the universal relation λ = Cμ, because the coefficient C differs for different diagrams. Nevertheless, the relation λ ~ μ is valid for any diagram according to the dimension reasoning. Hence, the MS scheme corresponds to the averaging of the vertex $\Gamma_4$ over the momenta with the weight function localized at the scale μ. From this point of view, the MS scheme can be treated as "physical" and the reasoning from Footnote 7 is applicable to it.

## ACKNOWLEDGMENTS

This work was supported by the Russian Foundation for Basic Research (project no. 06-02-17541).